\documentclass[traditabstract]{aa}

\def \obj{HE~0435-1223} 
\usepackage{graphicx}
\usepackage{txfonts}
\usepackage{pstricks}
\include{epsf}
\usepackage{natbib}
\bibpunct{(}{)}{;}{a}{}{,}

\def\0{\phantom0}

\def\ho{$H_{\rm 0}$}
\def\kms{km s$^{-1}$}
\def\kmsmpc{km s$^{-1}$ Mpc$^{-1}$}

\begin{document}

\title{COSMOGRAIL:  the COSmological MOnitoring of \\GRAvItational Lenses\thanks{Based on observations made with the 1.2m Euler Swiss Telescope, the 1.5m telescope of Maidanak Observatory in Uzbekistan, and  with the 1.2m Mercator Telescope, operated on the island of La Palma by the Flemish Community, at the Spanish Observatorio del Roque de los Muchachos of the Instituto de Astrof\'isica de Canarias. The NASA/ESA Hubble Space Telescope data was obtained from the data archive at the Space Telescope Science Institute, which is operated by AURA, the Association of Universities for Research in Astronomy, Inc., under NASA contract NAS-5-26555.}
\subtitle{IX.  Time delays, lens dynamics and baryonic fraction in \obj}}

\titlerunning{\obj: Time delays, dynamics and baryonic fraction of the lens}

\author{F. Courbin
\inst{1}
\and V. Chantry
\inst{2}
\thanks{Research Fellow, Belgian National Fund for Scientific Research (FNRS)}
\and Y. Revaz
\inst{1}
\and D. Sluse
\inst{3}
\thanks{Alexander von Humboldt fellow}
\and C. Faure
\inst{1}
\and M. Tewes
\inst{1}
\and E. Eulaers
\inst{2}
\and M. Koleva
\inst{4, 5, 6}
\and
I. Asfandiyarov
\inst{7}
\and S. Dye
\inst{8}
\and P. Magain 
\inst{2}
\and H. van Winckel
\inst{9}
\and J. Coles
\inst{10}
\and P. Saha
\inst{10}
\and M. Ibrahimov
\inst{7}
\and G. Meylan
\inst{1}
}

\institute{Laboratoire d'Astrophysique, Ecole Polytechnique F\' ed\' erale de Lausanne (EPFL), Observatoire de Sauverny, CH-1290 Versoix, Switzerland
\and
Institut d'Astrophysique et de G\' eophysique, Universit\' e de Li\`ege, All\'ee du 6 Ao\^ut, 17, 4000 Sart Tilman (Bat. B5C), Li\`ege 1, Belgium
\and
Astronomisches Rechen-Institut am Zentrum f\"ur Astronomie der Universit\"at Heidelberg,
M\"onchhofstrasse 12-14, 69120 Heidelberg, Germany
\and
Universit\'e Lyon 1, Villeurbanne, F-69622, France; CRAL, Observatoire de Lyon, St Genis Laval, F-69561, France; CNRS, UMR 5574
\and
Instituto de Astrof\'{\i}sica de Canarias, La Laguna, E-38200 Tenerife, Spain
\and
Departamento de Astrof\'{\i}sica, Universidad de La Laguna, E-38205 La
Laguna, Tenerife, Spain
\and
Ulugh Beg Astronomical Institute, Academy of Sciences, Tashkent, Uzbekistan
\and
Cardiff University, School of Physics and Astronomy,
Queens Buildings, The Parade, Cardiff, CF24 3AA, U.K.
\and
Instituut voor Sterrenkunde, Katholieke Universiteit Leuven, Celestijnenlaan 200B, 3001 Heverlee, Belgium
\and
Institute of Theoretical Physics, University of Z\"urich, Winterthurerstrasse 190, 8057 Z\"urich, Switzerland
}

\date{}
 
\abstract{We present accurate time delays for the quadruply imaged quasar \obj. 
The delays were measured from 575 independent photometric points obtained in the R-band between January 2004 and March 2010. With seven years of data, we clearly show that quasar image A is affected by strong microlensing variations and that the time delays are best expressed relative to quasar image B. We measured $\Delta t_{\mathrm{BC}} = 7.8\pm0.8$ days, $\Delta t_{\mathrm{BD}} = -6.5\pm0.7$ days and $\Delta t_{\mathrm{CD}} = -14.3\pm0.8$ days. 
We spacially deconvolved HST NICMOS2 F160W images to derive accurate astrometry of the quasar images and to infer the light profile of the lensing galaxy. We combined these images with a stellar population fitting of a deep VLT spectrum of the lensing galaxy to estimate the baryonic fraction, $f_b$, in the Einstein radius. 
We measured $f_b = 0.65^{+0.13}_{-0.10}$ if the lensing galaxy has a Salpeter IMF and $f_b = 0.45^{+0.04}_{-0.07}$ if it has a Kroupa IMF. The spectrum also allowed us to estimate the velocity dispersion of the lensing galaxy, $\sigma_{\rm ap}$ = 222 $\pm$ 34 \kms. 
We used $f_b$ and $\sigma_{\rm ap}$ to constrain an analytical model of the lensing galaxy composed of an Hernquist plus generalized NFW profile. We solved the Jeans equations numerically for the model and explored the parameter space under the additional requirement that the model must predict the correct astrometry for the quasar images. Given the current error bars on $f_b$ and $\sigma_{\rm ap}$, we did not constrain \ho\ yet with high accuracy, i.e., we found a broad range of models with $\chi^2 < 1$. However, narrowing this range is possible, provided a better velocity dispersion measurement becomes available. In addition, increasing the depth of the current HST imaging data of \obj\ will allow us to combine our constraints with lens reconstruction techniques that make use of the full Einstein ring that is visible in this object.

}
\maketitle
%

\begin{figure*} [t!]
\centering
\includegraphics[width=18.3cm]{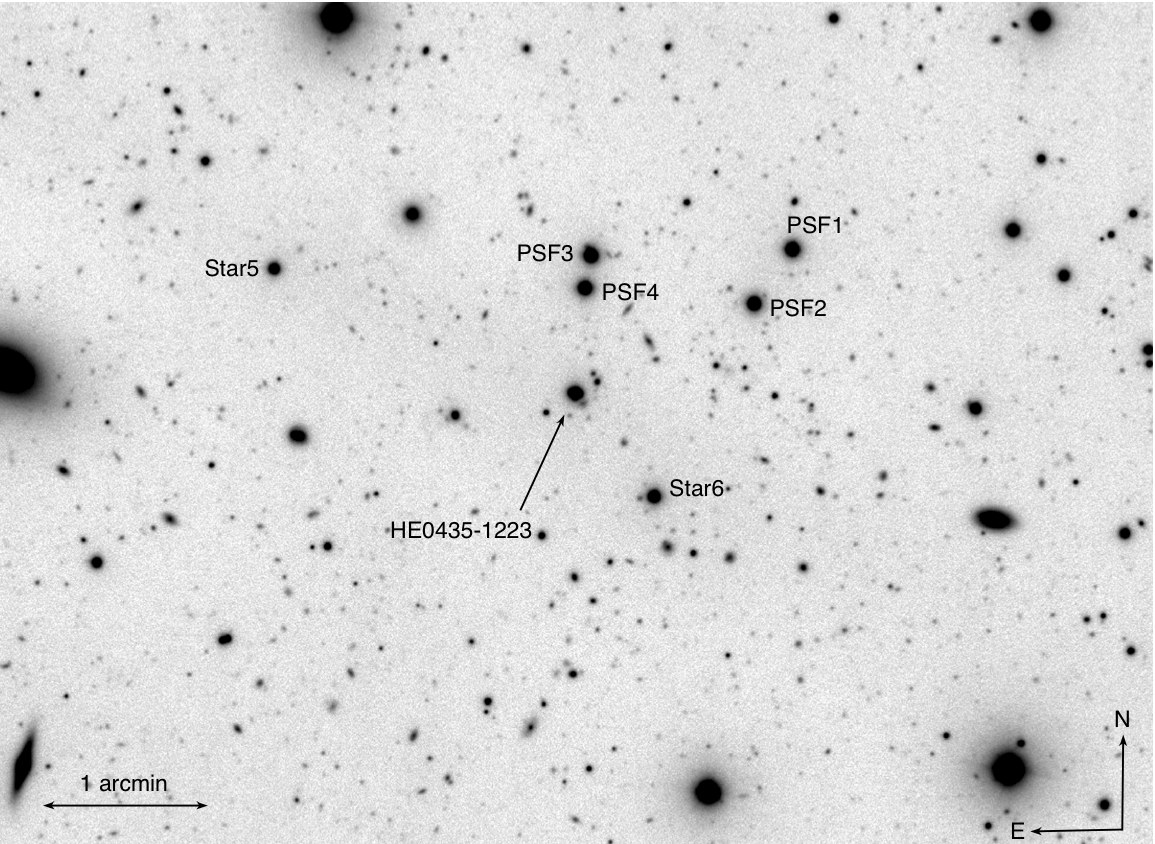}
\caption{Part of the field of view of the 1.2 m Swiss Euler telescope, with \obj\, visible in the centre. The four PSF stars used for deconvolution purposes and the two reference stars used to carry out the flux calibration are indicated.}
\label{field_Euler0435}
\end{figure*}

\section{Introduction}
\label{intro}

To determine the expansion rate of the Universe, \ho, accurately, it is important to scale the extragalactic distance ladder and
to measure $w(z)$, the redshift evolution of the dark energy equation of state 
parameter  \citep[e.g.,][]{Frieman08}.

Some of the most popular methods in use to measure \ho\, have recently been reviewed \citep{Freedman2010}. 
The huge observational and theoretical efforts invested in these measurements have led to random errors between
3\% and 10\%, depending on the methods. 
However, all these methods rely at some level on each other and are in the end all based on the same local standard candles. In addition, the accuracy required on \ho\, to measure $w(z)$ and H$(z)$ is of the order of 1\%  \citep{Hu2005, Riess2009, Riess2011}.
It is therefore of interest (i) to explore 
methods that are fully independent of any standard candle and (ii) to combine the different methods to further reduce
the current error bar on \ho.

Strong gravitational lensing of quasars and the so-called ``time delay method" in multiply imaged quasars 
is independent of the traditional standard candles \citep{Refsdal1964} and is based on well understood 
physics: general relativity.  
However, it requires to measure the time delays from long-term photometric monitoring of the many lensed
quasars, which has long been a serious observational limitation. While the attempts to measure accurate time delays have 
been numerous, only few quasars have been measured with an accuracy close to the percent 
\citep[e.g.,][]{Goiacoechea2002, Fassnacht2002, Cosmograil7}. In addition, the lens model necessary to convert time delays into \ho\, often remains poorly constrained and hampers a breaking of the degeneracies between the properties
of the lensing galaxy and \ho. In recent years, successful attempts have been made to constrain the lens model as much as possible to break these degeneracies 
\citep{Suyu2009, Suyu2010}. 

COSMOGRAIL, the COSmological MOnitoring of GRAvItational Lenses, aims both at measuring precise
time delays for a large sample of strongly lensed quasars, and at obtaining and using all necessary observations to constrain the lens models. The present paper describes the COSMOGRAIL results for
the quadruply imaged quasar \obj, using deep VLT spectra and deconvolved HST images.

\obj\, ($\alpha(2000): 04\, h\, 38\, min\, 14.9\, sec$; $\delta(2000)= -12^{\circ}17\arcmin 14\farcs4$) was discovered by \citet{Wisotzki2000} during the Hamburg/ESO Survey (HES) for bright quasars in the Southern Hemisphere. It was identified two years later as a quadruply imaged quasar by \citet{Wisotzki2002}. The redshift of the source is $\rm z_{\rm s}=1.689$ \citep{Wisotzki2000} and that of the lens is $\rm z_{\rm l}=0.4546 \pm 0.0002$ \citep{Morgan2005}. The quasar shows evidence for intrinsic variability, which makes it a good candidate for determining the time delays between the different images. The local environment of the lensing galaxy has been studied in detail by \citet{Morgan2005} using the Hubble Space Telescope (HST) Advanced Camera for Surveys (ACS) and by \citet{Momcheva2009}, who has found that the lensing galaxy lies in a group of at least 11 members. The velocity dispersion of this group is $\sigma\sim 496$ \kms.

Analytical lens models of \obj\, are given by \citet{Kochanek2006}, who have also measured time delays from
two years of optical monitoring: $\Delta t_{AD}=-14.37^{+0.75}_{-0.85}$ days, $\Delta t_{AB}=-8.00^{+0.73}_{-0.82}$ days, and $\Delta t_{AC}=-2.10^{+0.78}_{-0.71}$ days. For a fixed $H_{\rm 0} = 72\ \pm \ 7 \rm \ km \ s^{-1}Mpc^{-1}$ they found that the lensing galaxy must have a rising rotation curve at the position of the lensed images and a non-constant mass-to-light ratio. Moreover, high dark matter surface densities are required in the lens halo. New monitoring data of \citet{Blackburne2010} analysed using a physically motivated representation
of microlensing give time delays compatible with those of \citet{Kochanek2006}, although these authors
do not provide their measured values.

\begin{figure*} [t!]
\centering
\includegraphics[scale=0.7]{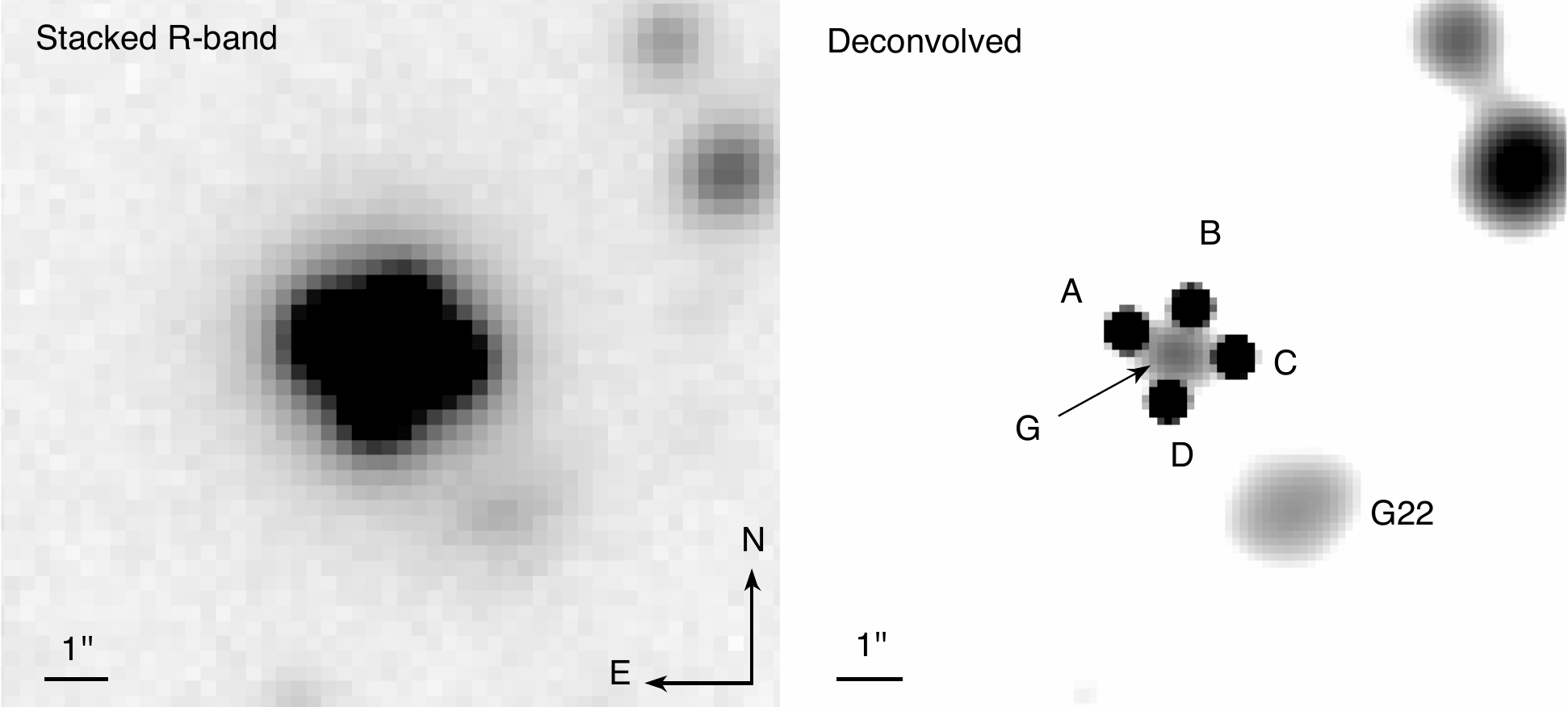}
\caption[Result from the simultaneous deconvolution of the ground-based frames]{Result from the simultaneous deconvolution of the ground-based frames. G is the lensing galaxy and G22 \citep{Morgan2005} is its closest neighbour on the plane of the sky. The grey scale in the deconvolved image is set to display all light level above $3 \times \sigma_{sky}$. The FWHM resolution of the deconvolved
image is 0\farcs34.}
\label{dec0435}
\end{figure*}

\begin{table*}[]
\centering 
\caption{Summary of the optical monitoring data. The temporal sampling is the mean number of days between two consecutive
observations.}
\begin{tabular}{lcccrrrcr}
\hline
Telescope & Camera & FoV & Pixel & Period of observation & \#obs. & Exp. time & Seeing & Sampling \\
\hline
\hline
Euler         & C2 & 11\arcmin $\times$ 11\arcmin & 0\farcs344 & Jan 2004 - Mar 2010 & 301 & 5 $\times$ 360s & 1\farcs37 & 6 days\\
Mercator   & MEROPE & 6.5\arcmin $\times$ 6.5\arcmin & 0\farcs190 &Sep 2004 - Dec 2008 & 104 & 5 $\times$ 360s& 1\farcs59 & 11 days\\
Maidanak & SITE & 8.9\arcmin $\times$ 3.5\arcmin & 0\farcs266 & Oct 2004 -  Jul 2006 & 26 & 10 $\times$ 180s & 1\farcs31 & 16 days\\
Maidanak & SI  & 18.1\arcmin $\times$ 18.1\arcmin & 0\farcs266 &  Aug 2006 -  Jan 2007 & 8 & 6 $\times$ 300s & 1\farcs31 & 16 days\\
SMARTS & ANDICAM & 10\arcmin $\times$ 10\arcmin & 0\farcs300 & Aug 2003 - Apr 2005 & 136 & 3 $\times$ 300s & $\le$ 1\farcs80 & 4 days\\
\hline
{\bf TOTAL} & - & - & - & Aug 2003 - Mar 2010 & 575 & 242.5 h & -  & 3.2 days\\
\hline
\end{tabular}
\label{data_sum0435}
\end{table*}

\section{Photometric monitoring}
\label{monitor0435}

\subsection{Optical imaging}

\obj\, was monitored during more than six years, from January 2004 to March 2010, through the R filter, using three 
different telescopes: the Swiss 1.2m Euler telescope located on the ESO La Silla site (Chile), the Belgian-Swiss 1.2m 
Mercator telescope located at the Roque de Los Muchachos Observatory, La Palma, Canary Island (Spain), and 
the 1.5m telescope located at the Maidanak Observatory (Uzbekistan). In addition we also use 136 epochs from 
the two-year long monitoring of \cite{Kochanek2006}, from August 2003 
to April 2005, obtained with the ANDICAM camera mounted on the 1.3m Small and Moderate Aperture Research 
Telescope System (SMARTS) located at the Cerro Tololo Inter-American Observatory (CTIO) in Chile. 
A summary of the observations is given in Table \ref{data_sum0435}.
Note that we used the published photometry for the SMARTS data, i.e., we did not reprocess the original data frames 
with our own photometric pipeline.

\begin{figure*} [ph!]
\centering
\includegraphics[scale=0.9]{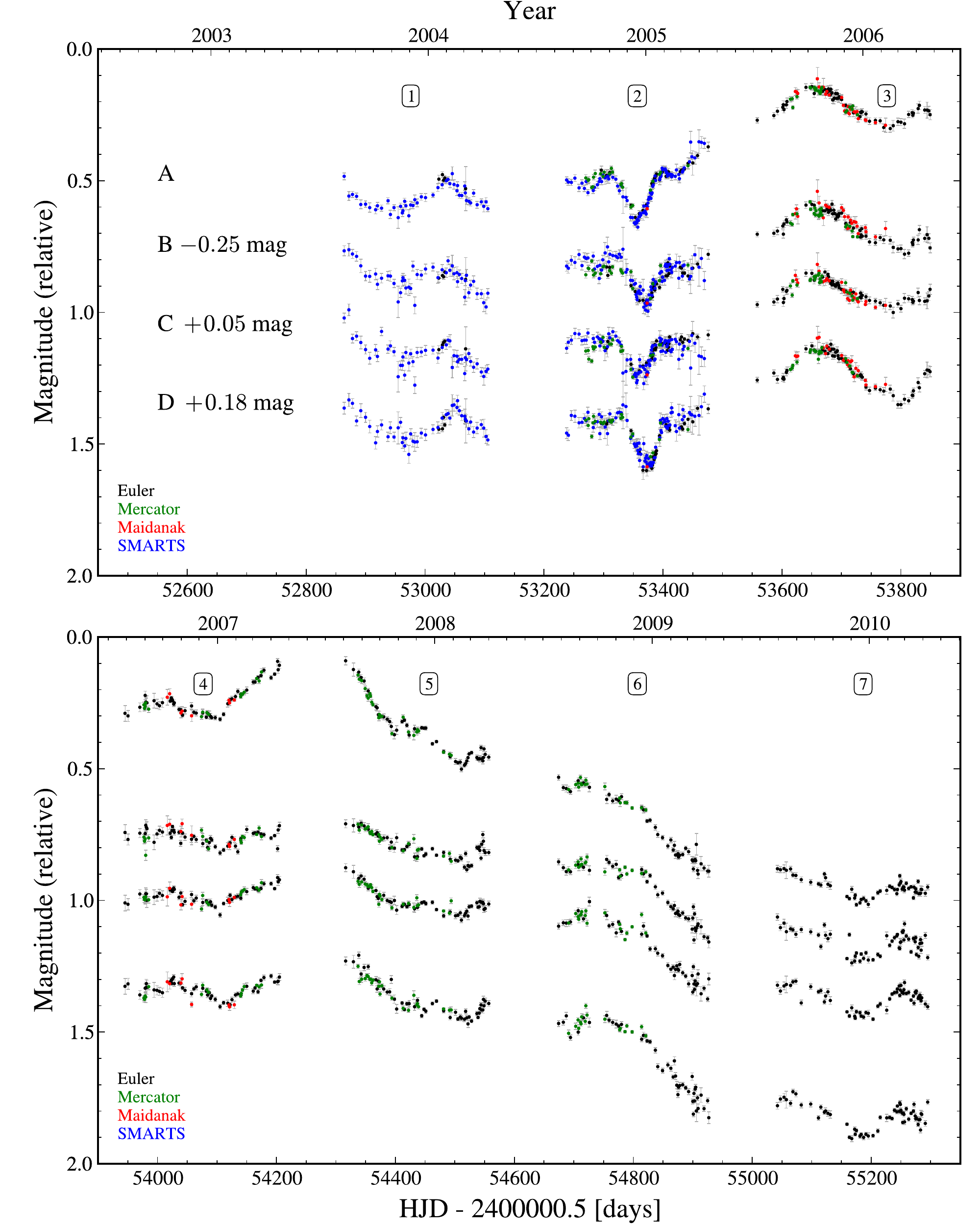}
\caption{R-band light curves of the four lensed images of \obj\, from December 2003 to April 2010. The magnitudes are  given in relative units as a function of the Heliocentric Julian Day (HJD), along with their total $1 \sigma$ error bars. 
}

\label{lightcurves0435}
\end{figure*}

\begin{figure*} [t!]
\centering
\includegraphics[scale=0.353]{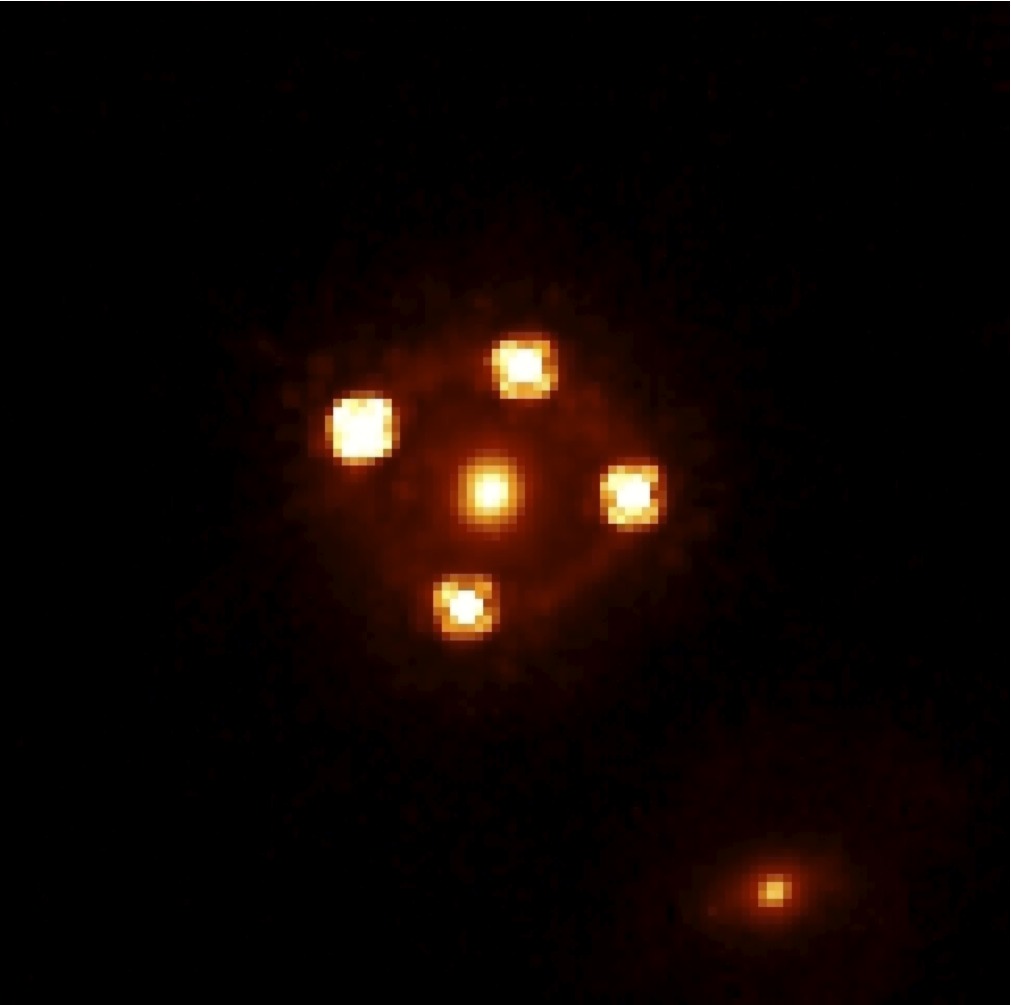}
\includegraphics[scale=0.353]{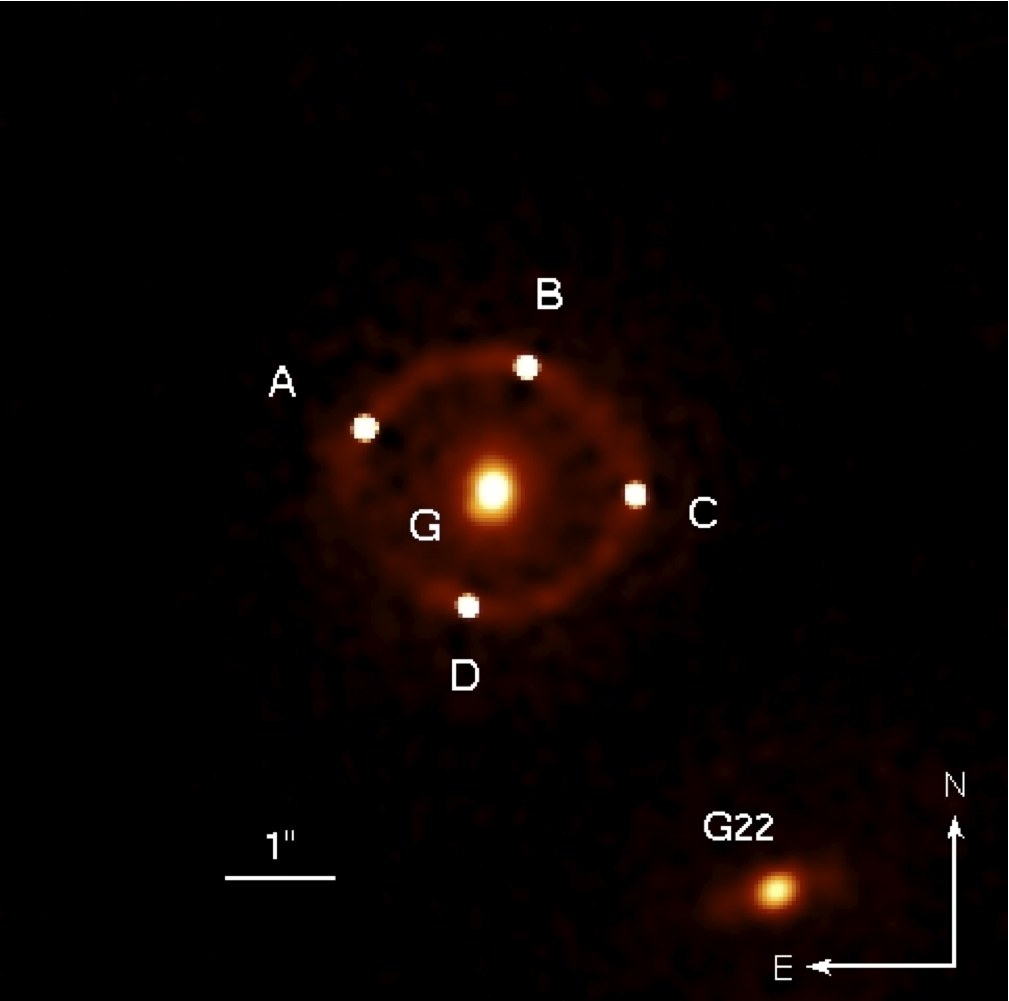}
\includegraphics[scale=0.353]{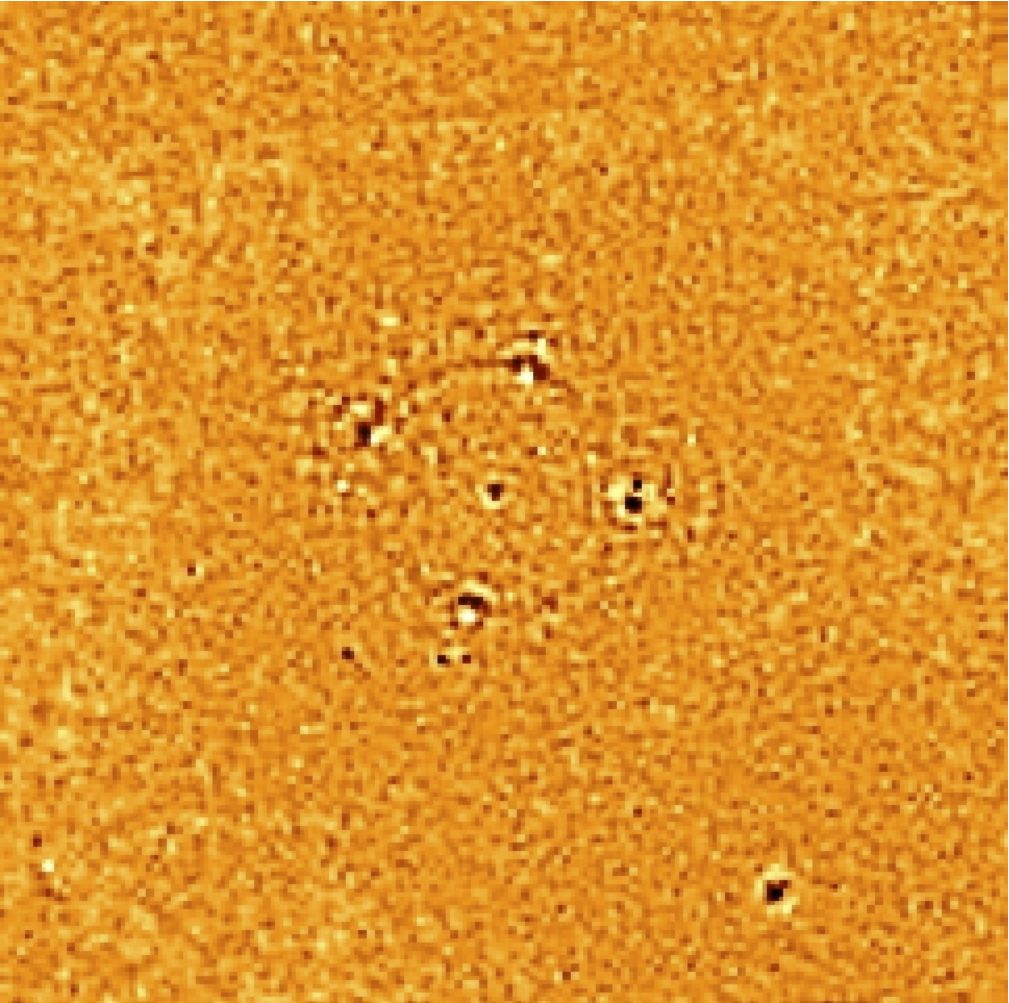}
\caption{\textit{Left}: Combination of the four original HST/NIC2 F160W frames of \obj. The field of view is 9 $\times$ 9 arcsec. 
\textit{Middle}: Deconvolved image, where the lensing galaxy is modelled as a de Vaucouleurs profile (see text). The nearest 
galaxy on the plane of the sky, G22, is also indicated. \textit{Right}: Residual map in units of the noise.  
The colour scale ranges from -4$\sigma$ (white) to +4$\sigma$ (black).}
\label{dec_Nic0435}
\end{figure*}

\begin{table*}
\centering 
\caption{Relative astrometry of \obj as derived from the simultaneous deconvolution of all NIC2 frames.  
The 1$\sigma$ error bars are the internal errors after deconvolution. Additional 2-mas systematic errors must be added
to these (see text). The magnitudes are in the Vega system. 
For comparison we show the results from \citet{Morgan2005} using HST/ACS images and from \citet{Kochanek2006} 
using HST/NIC2 images.}
\begin{tabular}{cccccccc}
\hline
\hline
& \multicolumn{3}{c}{This work} & \multicolumn{2}{c}{\citet{Morgan2005}} & \multicolumn{2}{c}{\citet{Kochanek2006}} \\

ID & $\Delta \alpha$ (\arcsec) & $\Delta \delta$ (\arcsec) & Mag (F160W) & $\Delta \alpha$ (\arcsec) & $\Delta \delta$ (\arcsec) & $\Delta \alpha$ (\arcsec) & $\Delta \delta$ (\arcsec) \\ 
\hline
A & 0.        & 0.        &  17.20 $\pm$ 0.01 &  0.        & 0.        
	& 0.         & 0.         \\
B & -1.4743 $\pm$ 0.0004 & +0.5518 $\pm$ 0.0006 &  17.69 $\pm$ 0.01  & -1.477  $\pm$ 0.002 & +0.553 $\pm$ 0.002
	& -1.476  $\pm$ 0.003 & +0.553 $\pm$ 0.001  \\
C & -2.4664 $\pm$ 0.0003 & -0.6022 $\pm$ 0.0013 & 17.69 $\pm$ 0.02 & -2.469 $\pm$ 0.002 & -0.603 $\pm$ 0.002
	& -2.467 $\pm$ 0.002 & -0.603 $\pm$ 0.004  \\
D & -0.9378 $\pm$ 0.0005 & -1.6160 $\pm$ 0.0006 & 17.95 $\pm$ 0.01& -0.938 $\pm$ 0.002 & -1.615 $\pm$ 0.002
	& -0.939  $\pm$ 0.002 & -1.614 $\pm$ 0.001 \\
G & -1.1706 $\pm$ 0.0030  & -0.5665 $\pm$ 0.0004 & 16.20 $\pm$ 0.12 & -1.169 $\pm$ 0.002 & -0.572 $\pm$ 0.002 
&  -1.165 $\pm$ 0.002 & -0.573 $\pm$ 0.002 \\
\hline
\end{tabular}
\label{astrom0435}
\end{table*}

\subsection{Image processing and deconvolution photometry}

The data from the three telescopes used by the COSMOGRAIL collaboration are analysed with the semi-automated reduction pipeline described in \cite{Cosmograil5}. The main challenge was that we had to assemble data from different telescopes: each camera has a different size, resolution and orientation on the plane of sky. 

The pre-reduction for each observing epoch consists of flat-fielding using master sky-flats. The Euler image with the best 
seeing was taken as the reference frame to register all other frames. 
This reference frame was taken on the night of November 11, 2005 and has a seeing of 0\farcs82. Two reference stars were then chosen in the field of view (Fig. \ref{field_Euler0435}) to compute the geometrical transformations between the images. This transformation involves a spatial scaling and a rotation. The reference stars were also used to compute the relative photometric scaling between the frames taken at 
different epochs. Eventually, the L.A. Cosmic algorithm \citep{lacos} was applied separately to every frame to remove cosmic rays. All images were checked visually to make sure that no pixel was removed inappropriately, especially in the frames with good seeing.

The photometric measurements were carried out using ``deconvolution photometry" with the 
MCS deconvolution algorithm \citep{MCS98}. This software has been successfully applied to
a variety of astrophysical problems ranging from gravitationally lensed quasars \citep[e.g.,][]{Burud2000,Burud2002} to the study of quasar host galaxies \citep[e.g.,][]{Letawe2008}, or  to the search for extrasolar planets using the transit technique \citep[e.g.,][]{Gillon2007,Gillon2007a}. Image deconvolution requires accurate knowledge of the instrumental and atmospheric point spread function (PSF). The latter was computed for each frame from the four stars labelled PSF 1-4 in Fig.~\ref{field_Euler0435}. 
These stars are from 0 to 1 magnitude brighter than the quasar images in \obj\, and are located within 2\arcmin\, from
the centre of the field, which minimizes PSF distortions.

Because it does not attempt to achieve an infinitely high spatial resolution, the MCS algorithm produces deconvolved images that are always compatible with the sampling theorem. This avoids deconvolution artefacts and allowed us to carry out accurate photometry over the entire field of view. Moreover, the deconvolved image was computed as the sum of extended numerical structures and of analytical point sources whose shape is chosen to be 
symmetrical Gaussians. In the case of gravitationally lensed quasars, the numerical channel of this 
decomposition contains the lensing galaxy. The photometry and astrometry of the quasar images were returned as a list of intensities and positions of Gaussian deconvolved profiles. Finally, the deconvolved image can be computed on a grid
 of pixels of arbitrary size. In the present work, the pixel size in the  deconvolved frames is half the pixel size of the Euler
data, i.e., 0\farcs172. The spatial resolution in the deconvolved frames is two pixels full-width-at-half-maximum
(FWHM), i.e., 0\farcs35. 
 
With the MCS software, dithered images of a given target can be ``simultaneously deconvolved" and combined into 
a single deep and sharp frame that matches the whole dataset at once, given the  PSFs and the noise maps of the
individual frames. 
In doing this, the intensities of the point sources are allowed to vary from 
one frame to the next while the smooth background, which includes the lensing galaxy, is held constant in all frames. 
The result of the process is shown in Fig. \ref{dec0435}, where the point sources are labelled as in \cite{Wisotzki2002}. 
Prior information on the object to be deconvolved can be used to achieve the best possible results. In the case 
of \obj\, the relative positions of the point sources are fixed to the HST astrometry obtained in Sect. \ref{Nicmos0435}.

Fig. \ref{lightcurves0435} shows the deconvolution light curves obtained for each quasar image of \obj, where
the $1 \sigma$ error bars account both for the statistical and systematic errors. The statistical part of the error 
was taken as the dispersion between the photometric points taken during each night. The systematic errors were estimated by carrying out the simultaneous deconvolution of reference stars in the vicinity of \obj.

Finally, a small scaling factor was applied to the light curves of all telescopes, including the published light curves  of \citet{Kochanek2006}, to match the Euler photometry. These shifts are all smaller than 0.03 mag.

\begin{table}[t!]
\centering 
\caption{Shape parameters for the lensing galaxy in \obj. The position angle (PA)
is measured positive east of north. The $1 \sigma$ error bars (internal errors) 
are given in parenthesis.}
\begin{tabular}{ccccc}
\hline
\hline
PA ($^{\circ}$) & Ellipticity & $a_{eff}$ (\arcsec) & $b_{eff}$ (\arcsec) & $r_{eff}$ (\arcsec) \\ 
\hline
174.8 (1.7) & 0.09 (0.01) & 1.57 (0.09) & 1.43 (0.08) & 1.50 (0.08)  \\
\hline
\end{tabular}
\label{gal0435}
\end{table}

\section{HST NICMOS2 imaging}
\label{Nicmos0435}

We used deep near-IR HST images of \obj\, to derive the best possible relative astrometry
between the quasar images and the lensing galaxy and to constrain the light distribution in the lensing
galaxy. The data are part of the CASTLES project (Cfa-Arizona Space Telescope LEns Survey) 
and were acquired in October 2004 (PI: C.S. Kochanek) with the camera 2 of NICMOS, the 
Near-Infrared Camera and Multi-Object Spectrometer. They consist of four dithered frames taken through the
F160W filter (H-band) in the MULTIACCUM mode with 19 samples and calibrated by CALNICA, 
the HST image reduction pipeline. The total exposure time amounts to approximately 
44 minutes and the pixel scale is 0\farcs075652.

The MCS deconvolution algorithm was used to combine the four NIC2 frames into a deep sharp IR image. 
We followed the iterative technique described in \cite{Chantry2010} and \citet{chantry2007}, which allowed us to build a 
PSF in the absence of a stellar image in the field of view. The method can be summarised as follows. First, we estimated the PSF using {\tt Tiny Tim} software \citep{Tinytim} and carried out the simultaneous deconvolution of the 
four F160W frames using a modified version of the MCS software  \citep{PSFsimult}. This produces a first 
approximation of the extended channel of the deconvolved image, i.e., the lensing galaxy and the lensed 
quasar host galaxy. We reconvolved the latter by the PSF and subtracted it from the original data. 
A new estimate of the PSF was built on the new image that now contained only the quasar images.
The process was repeated until the residual image was satisfactory \citep[for more details see][]{Chantry2010}.
Fig. \ref{dec_Nic0435} shows the result. In this image the pixel size is half that of the original data and the
resolution 0\farcs075 (FWHM), unveiling an almost full Einstein ring.

In the final deconvolved image, the 
lensing galaxy was modelled analytically rather than numerically to minimise the number of degrees of 
freedom. We found that the best-fit profile is an elliptical de Vaucouleurs with the parameters 
as given in Table~\ref{gal0435}. The astrometry of the quasar images relative to the lensing galaxy,
corrected for the known distortions of the NIC2 camera and for the difference of pixel scale between 
the $x$ and $y$ directions is summarised  in Table~\ref{astrom0435}. Based on our previous 
work using deconvolution of NICMOS images \citep{chantry2007}, we estimate that the total error bars, accounting 
for residual correction of the distortions amounts to 2 mas. Our results agree well
with previous measurements from HST/ACS \citep{Morgan2005} or HST/NIC2 imaging \citep{Kochanek2006},
also shown for comparison in Table~\ref{astrom0435}.

\begin{figure*} [t!]
\centering
\includegraphics[width=1.0\textwidth, height=14cm]{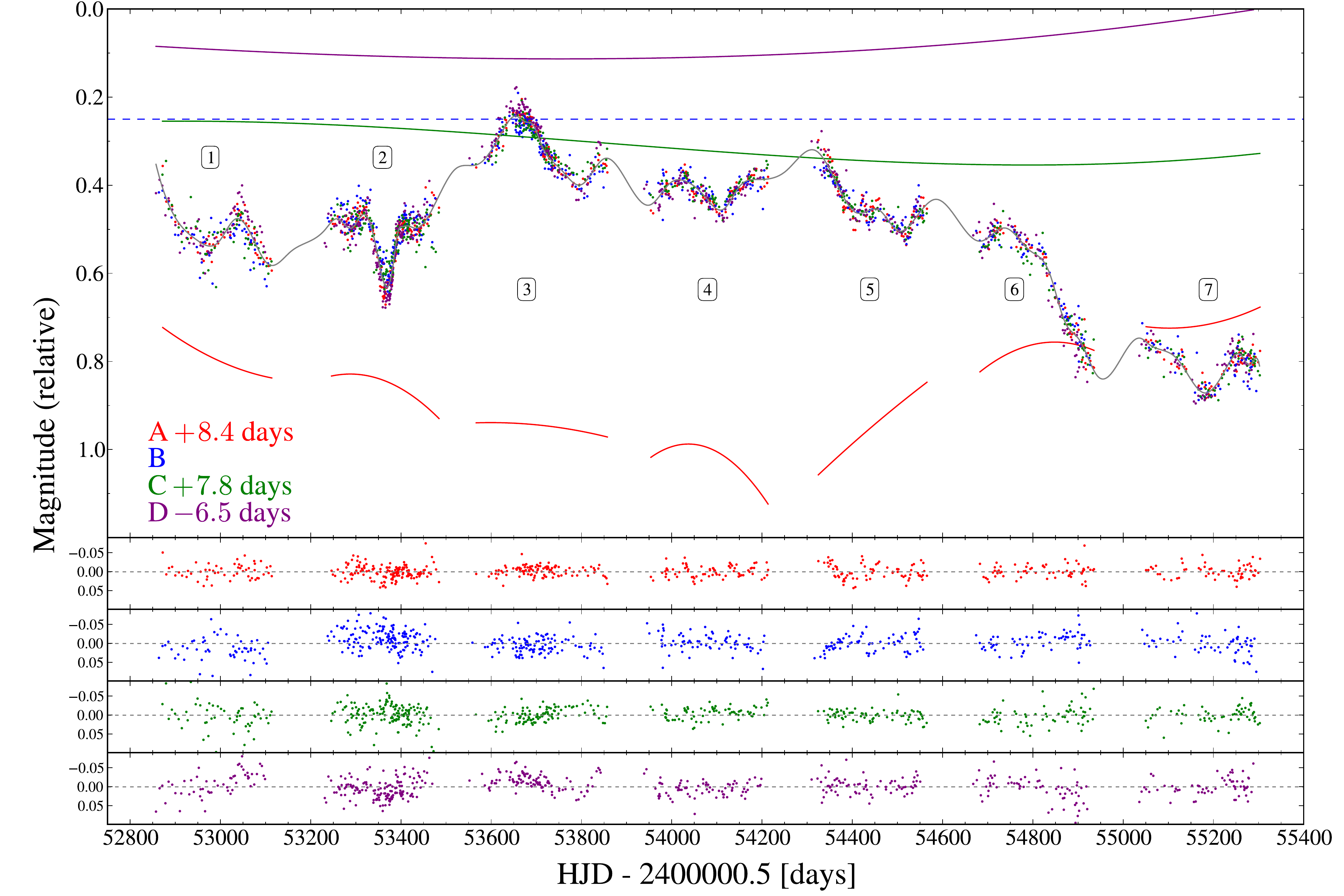}
\caption{Lightcurves obtained with all four telescopes and shifted by time delays of 
$\Delta t_{\mathrm{BA}} = 8.4$ days,  $\Delta t_{\mathrm{BC}} = 7.8$ days 
and $\Delta t_{\mathrm{BD}} = -6.5$ days. 
The relative microlensing representations applied on curves A, C and D are shown as continuous curves with respect to the dashed blue line (see text). A fifth-order polynomial was used over the seven seasons to model microlensing on the quasar images C and D, while seven independent third-order polynomials were used for image A.
The lower panels show residuals obtained by subtracting a simultaneous spline fit (grey) from the light curves.
}
\label{microlensing}
\end{figure*}

\section{Time delay measurement}
\label{delay0435}

\subsection{Curve shifting method}

Our method to measure the time delays is based on the dispersion technique of \citet{Pelt96}: the light curves
are shifted in time and in magnitude to minimise a global dispersion function. In addition, the light curves are distorted on long time 
scales to account for slow microlensing variations. This was made by adding low-order polynomials to either the full curves 
or to specific observing seasons. 

\citet{Pelt96} has defined several dispersion statistics between pairs of light curves. We implemented a dispersion estimate similar to $D_{3}^{2}$ \citep[see eq. 8 of][]{Pelt96},  
which performed a linear interpolation between points of one of the curves over a maximum range of 30 days.
In the case of four light curves, we defined a total dispersion that is the sum of the dispersions
computed using the 12 possible permutations of two curves among four. Each pair was considered twice so to avoid the arbitrary 
choice of a reference light curve. The photometric error bars were taken into account to weight the influence 
of the data points in the dispersion.
We then minimised the total dispersion by modifying the time delays and  the microlensing polynomials. 

\subsection{Microlensing and influence on the time delay}

Simulated light curves that mimic the observed data were used 
to estimate the robustness of the method. The error bars on the time delays were calculated using Monte Carlo simulations, i.e., 
redistributing the magnitudes of the data points according to their photometric error bars. The width of the 
resulting time delay distributions gives us the 1$\sigma$ error bars. 

Because of microlensing we do not have access to the intrinsic variations of the quasar. We represent microlensing in three of the light curves as a relative variation with respect to the fourth light curve, taken
as a reference. We tested each of the four light curves in turn as a reference and kept the one that
minimised the residual microlensing variations to be modelled in the 3 others. This is best verified
with component B as a reference.

With B as a reference light curve, we note that microlensing in C and D remains smooth and can therefore 
be modelled with a low-order polynomial drawn over the full length of the monitoring. However, A contains 
higher frequency variations that need to be accounted for in each season individually, as illustrated in 
Fig.~\ref{microlensing}. In doing this, we obtain fairly good fits to the light curves, as shown in the residual signal.
To quantify the quality of these residuals, we applied the so-called {\it one-sample runs test of randomness}, a statistical test to estimate whether successive rea\-lizations of a random variable are independent or not. In practice the test was applied to a sequence of residuals to decide whether a model is a good representation of the data. For most seasons in our curves the number of runs was between $1 \sigma$ and $3 \sigma$ lower than the value expected for independent random residuals. Thus, although our microlensing model is  not fully representative of the real signal, the deviations from the data points remain small. 

We tested the robustness of our curve-shifting method in several ways. First, we modelled the microlensing variations using polynomial fits of different orders. Second, we fitted these polynomials either across each individual season or across 
groups of seasons. Finally, we masked the seasons with the worst residual signal (Fig.~\ref{microlensing}). 
All these changes had only a negligible impact on the time delay measurements.
We note that this is not the case when considering only two or three seasons of data, which shows the importance
of a long-term monitoring with good temporal sampling.

\subsection{Final results}

Our results are summarised in Table~\ref{TD_0435} and are compared with the previous measurements of 
\citet{Kochanek2006}, who have used pure a polynomial fit to the light curves and two seasons of monitoring. 
Using the same data but with our modified  
dispersion technique, we obtained very similar time delays as \citet{Kochanek2006}, but larger
error bars. We prefer keeping a minimum possible
number of degrees of freedom (e.g., in the polynomial order used to represent microlensing), 
in accordance with the Occam's razor principle, even to the cost of apparently 
larger formal error bars. 

We also note that \citet{Kochanek2006} give their time delays with respect to A, which, with seven seasons of data, turns out to
be the most affected by microlensing. As a consequence the error bars on these time delays are dominated
by residual microlensing rather than by statistical errors. 
The time delays used in the rest of our analysis are therefore measured relative to B.

Finally, we used the mean values of our microlensing corrections to estimate the macrolensing R-band flux ratios 
between the four quasar images, assuming that no long-term micro\-lensing affects 
the data. We found $m_B - m_A = \Delta m_{BA} = 0.62 \pm 0.04$, $\Delta m_{BC} = 0.05 \pm 0.01$, and
$\Delta m_{BD} = -0.16 \pm 0.01$, which is well compatible with the ratios measured at seven wavelengths by 
\citet{Mosquera2010}. However, these authors report significant wavelength dependence of the image flux ratios, which
led us not to use flux ratios as a constraint in the lens models.

\begin{table*}
\caption{Time delays for \obj, with the same arrival order convention as \citet{Kochanek2006}, i.e., D arrives last.}
\label{TD_0435}
\begin{center}
\begin{tabular}{llcccccc}
\hline\hline
Data & Method & $\Delta t_{\mathrm{AB}}$ & $\Delta t_{\mathrm{AC}}$ & $\Delta t_{\mathrm{AD}}$ & $\Delta t_{\mathrm{BC}}$ & $\Delta t_{\mathrm{BD}}$ & $\Delta t_{\mathrm{CD}}$  \\
\hline

SMARTS (seasons 1 and 2) & Kochanek (2006) & $-8.0 \pm 0.8$ & $-2.1 \pm 0.8$ & $-14.4 \pm 0.8$ &  &  &  \\

SMARTS (seasons 1 and 2) & dispersion &  $-8.8 \pm 2.4$ & $-2.0 \pm 2.7$ & $-14.7 \pm 2.0$ &  $6.8 \pm 2.7$ & $-5.9 \pm 1.7$ & $-12.7 \pm 2.5$  \\ 

COSMOGRAIL (all seasons) & dispersion & $-8.4 \pm 2.1$ & $-0.6 \pm 2.3$ & $-14.9 \pm 2.1$ &  $7.8 \pm 0.8$ & $-6.5 \pm 0.7$ & $-14.3 \pm 0.8$  \\ 

\hline
\end{tabular}
\end{center}
\end{table*}


\begin{table*}[ht!]
\begin{center}
\caption{Model parameters of the lens potential well. For the variables used as parameters
the ranges of values used are given between brackets.}
\label{tab:nbody}
\begin{tabular}{c c l}
\hline
\hline
$R_{eff}$ 		& $8.44\,\rm{kpc}$ 		& radius containing 50\% of the observed light\\
$R_{E}$   		& $6.66\,\rm{kpc}$ 		& Einstein radius \\
$f_b$			& parameter			& baryonic fraction in the Einstein radius $[0.05 - 0.5]$ \\
\hline
$r_\star$   		& $5.3\,\rm{kpc}$ 		& stellar component scaling radius\\
$M_{\star}$ 		& $M_{\rm{h}} f_b/(1-f_b) $	& stellar total mass\\
$r_{\star,\rm{max}}$ 	& $20\,r_\star$ 		& stellar truncation radius\\
\hline
$\gamma_{DM}$			& parameter			& dark matter inner slope $[0 - 2]$ \\
$r_s$                   & parameter			& dark matter scaling radius $[1,2,4,8] \times R_{E}$\\
$M_{\rm{h}}$ 		& parameter			& dark matter total mass $[4.8 \times 10^{11} - 9.1 \times 10^{12}] \rm{M_\odot}$\\
$r_{\rm{s},\rm{max}}$   & $10\,r_{\rm{s}}$		& dark matter truncation radius\\
\hline
\end{tabular}
\end{center}
\end{table*}

\section{Constraining the mass profile of the lensing galaxy}

The goal of the present section is to constrain the radial mass profile of the lensing galaxy as much as possible, which is the main source of uncertainty on the determination of \ho\ with the time delay method. 
One way of doing this is to use the information contained in the Einstein
ring that is formed by the host galaxy of the lensed quasar \citep{Suyu2010, Suyu2009, Warren2003}. 
This works well when a 
prominent Einstein ring is visible. Unfortunately, given the depth of the current HST images of \obj,
the radial extent of the ring is too small to efficiently apply this technique. We propose instead
to use information on the dynamics and on the stellar mass of the lens, using
 deep optical spectroscopy.
 
\subsection{Stellar population and velocity dispersion of the lensing galaxy}
\label{stelpop}
\label{addconst}

A deep VLT spectrum of the lensing galaxy is available from \citet{Cosmograil3}. While the spectrum was originally 
used to measure the redshift of the galaxy, it turns out to be deep enough ($<S/N> \sim 20$) to measure the stellar
velocity dispersion and the mass-to-light ratio.

We analysed the data using full spectrum fitting with the ULySS package \citep{Koleva2009}. The method consists in fitting spectra against a grid of stellar population models convolved by a line-of-sight velocity distribution. A single minimisation allowed us to determine the population parameters
(age and metallicity) and the kinematics (redshift and velocity dispersion). The stellar mass-to-light ratio was derived from the age and metallicity for the considered model. The simultaneous fit of the kinematical and stellar population parameters reduced the degeneracies 
between age, metallicity and velocity dispersion \citep{kolevaAN}.

We performed the spectral fit using the PEGASE-HR
single stellar population models \citep[SSP,][]{LeBorgne2004}, where the observed flux, 
$F_\lambda$, is modelled as follows:

\begin{eqnarray}
F_\lambda & = & P_{n}(\lambda) \times  [L(v_{sys},\sigma) \otimes S(t, [Fe/H], \lambda)] \nonumber \\ 
 &  + & Q_{m}(\lambda). 
\end{eqnarray}

The models were built using the Elodie.3.1 \citep{Prugniel2001,Prugniel2007} spectral library and the Kroupa 
\citep{Kroupa2001} and Salpeter \citep{Salpeter1955} initial mass functions (IMF). 
$L(v_{sys},\sigma)$ is a Gaussian function of the systematic velocity, $v_{sys}$, and of the velocity 
dispersion, $\sigma$. $S(t, [Fe/H], \lambda)$ is the model for the SSP and depends on age and metallicity. 
$P_n$ is a polynomial of degree $n$, 
which models any residual uncertainty in the flux calibration and extinction correction. In addition, the quasar spectrum might not
be perfectly subtracted from the galaxy. To mimic this effect, we included an additive polynomial, $Q_m$. 
The order of the additive and multiplicative polynomials is the minimum required to provide an acceptable $\chi^2$, i.e., in our case $(n=10,p=1)$.
 
We obtained SSP-equivalent ages and metallicities of $t\sim3$\,Gyr and $[Fe/H]\sim0.0$\,dex respectively. 
The corresponding rest-frame B-band stellar mass-to-light ratio is $M_\star / L_B = 3.2^{+0.3}_{-0.5}\ M_\odot / L_{\odot, B}$ 
using a Kroupa IMF and $M_\star / L_B = 4.6^{+0.9}_{-0.7}\ M_\odot / L_{\odot, B}$ using a Salpeter IMF. 
The uncertainties in the age and metallicity were estimated via Monte Carlo simulations and 
propagated in the error in $M_\star / L_B$.

To compute the physical velocity dispersion we subtracted quadratically the instrumental broadening from the measured
profile, neglecting the dispersion of the models since they are based on high-resolution templates. 
The instrumental broadening was measured both from the PSF stars used to carry out the spatial deblending of the
spectrum \citep{Cosmograil3} and from the lamp spectra. We obtained the rest-frame physical stellar velocity 
dispersion of the lensing galaxy: $\sigma_{\rm ap}$ = 222 $\pm$ 34 \kms\ in an aperture of 1\arcsec, i.e., 5.7 kpc.

\subsection{Numerical integration of the Jeans equations}
\label{nbs}

In this section we model the lensing galaxy using a 3D spherical potential well formed of two components,
one for the stellar part of the mass and one for the dark matter halo.  We then  
perform a numerical integration of the Jeans equations in 3D to predict a theoretical velocity dispersion 
and a total mass for the model. 
The assumption of spherical symmetry is sufficient for our purpose, as illustrated by the study of  the
lensed quasar MG~2016+112, where \citet{Koopmans2002a} introduced an anisotropy parameter and showed that 
it has almost no influence on the inferred mass slope.

The luminous component of the model is a Hernquist profile \citep{Hernquist1990}: 
\begin{equation}
\rho_{\star}(r)=\frac{\rho_{\star}(0)}{ \left( r/r_\star \right)  \left( 1 +r/r_\star \right)^3},
\label{Hernquist}
\end{equation} 
where $\rho_{\star}(0)$ is the central density and $r_\star$ is a scale radius chosen so that 
the integrated mass in a cylinder of radius $R_{\rm{eff}}$ (effective radius) is equal to half
the total stellar mass $M_{\star}$. The profile has a maximum radius of $r_{\star,\rm{max}}=20\,r_\star$. 

The dark matter halo is modelled as a generalised Navarro, Frenk \& White (NFW) profile \citep{NFW1996}:

\begin{equation} \rho_h(r)=\frac{\rho_{h}(0)} {(r/r_s)^{\gamma_{DM}} (1+(r/r_s)^2)^{(3 - \gamma_{DM})/2} },
\label{halo}
\end{equation} 
where $\gamma_{DM}$ is the inner slope of the profile, $r_s$ is the scaling radius and
$\rho_{h}(0)$ is the central mass density. 
For $\gamma_{DM}=1$ the model closely follows the standard NFW profile. Its total mass, 
$M_{\rm{h}}$, is given in the truncation radius 
$r_{\rm{h},\rm{max}}=10\,r_s$.

Following the usual convention, the integrated stellar and dark matter masses are related by the 
baryonic fraction $f_b$,
\begin{equation}
f_b = \frac{M_{\star} }{ M_{\rm{h}}+M_{\star} }.
\end{equation} 

Because all integrations in this work were carried out numerically, $f_b$ can easily be computed in any aperture. We chose to compute it in the Einstein radius, which is also where lensing gives the most accurate mass measurement.
The velocity dispersion of the stellar component was computed by solving the second moment of the Jeans equation in spherical coordinates \citep{Binney1998}. The velocity dispersion is then
\begin{equation}
         \sigma_{\star}^2(r) = \frac{1}{\rho_{\star}(r)}\int_r^\infty\! dr' \,\rho_{\star}(r')\, \partial_{r'} \Phi(r'),
        \label{sr_sph}
\end{equation}

where $\Phi(r)$ is the total gravitational potential. Equation~(\ref{sr_sph}) is solved numerically as follows.
First, the density of each mass component (stars and dark halo) is sampled
by N particles, using a Monte-Carlo method. Second, the potential is computed using a \emph{treecode} 
method \citep{barnes86}. For self-consistency, the density is computed by 
binning the particles in spherical shells. Then, a velocity is allocated to each particle at a distance $r$ from the galaxy 
center, following a Gaussian distribution of variance 
$\sigma_{\star}(r)$. Finally, the
velocity dispersion of the model, $\sigma_{\rm Jeans}$, is computed numerically, by integrating all the particle 
velocities in an aperture
that matches exactly the slit used to carry out the observations.

The advantages of the numerical representation of the lens models are multiple. 
They allow us (i) to compute velocity dispersions for any 
combination of density profiles, even non-parametric ones, (ii) to account for 
the truncation radius of the halo, (iii) to test the dynamical 
equilibrium of the system by evolving the model with time. 
Computing the velocity dispersion as well as the total mass in a cylinder along the line of sight of the observer
is then straightforward.

In all calcultations we fixed the effective radius to the one observed for the lensing galaxy, i.e., 
$R_{\rm eff} = 8.44$ kpc. The remaining parameter space to explore is composed of 
the halo scale radius, $r_s$, 
the slope of the profile, $\gamma_{DM}$, and the baryonic fraction, $f_b$, within the Einstein radius.

\begin{figure}[t!]
\centering
\includegraphics[width=91mm]{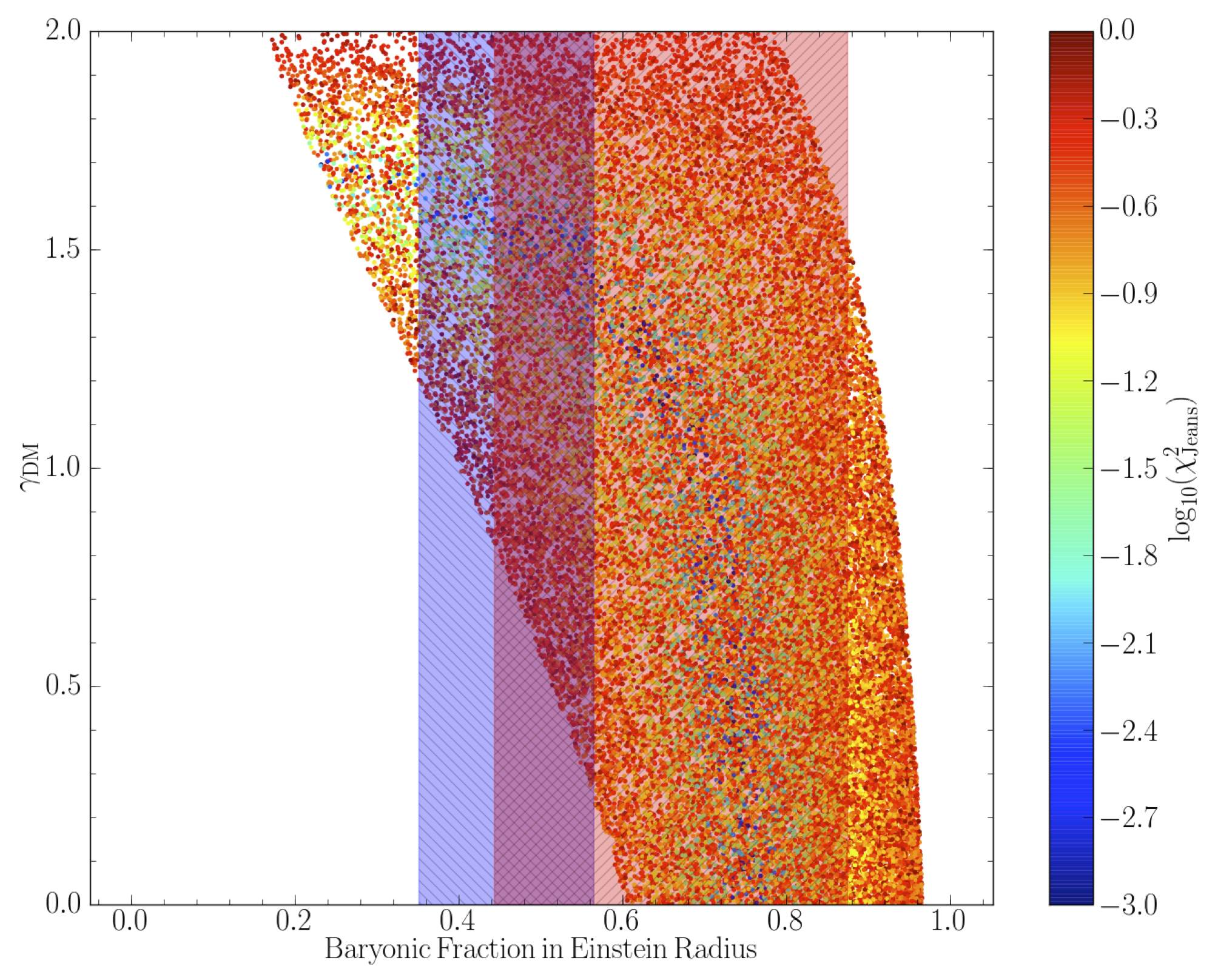}
\caption{Radial slope, $\gamma_{DM}$, of the dark matter halo as a function of 
the baryonic fraction in the Einstein radius, $R_E$. The colour code represents the value of 
$\chi^2_{\rm Jeans}$ for all acceptable models, i.e., $\chi^2_{\rm Jeans}<1$ (see text; eq.~\ref{eq:chijeans}). 
The observational limits drawn on $f_b$ using our spectrum of the lensing
galaxy are indicated as shaded vertical regions. The blue region corresponds to a Kroupa IMF and the red regions
corresponds to a Salpeter IMF.
In this figure, all models have $r_s = 4 \times R_E = 25$ kpc.}
\label{fb-gamma}
\end{figure}

\subsection{Applying the dynamical and stellar population constraints}

Using our HST photometry, we measured the total rest-frame B-band galaxy luminosity by converting
its total H-band magnitude, $m_{F160W}=16.20$. Using a k-correction of 1.148, and a galactic
extinction\footnote{A galactic extinction calculator is available at this address:
http://nedwww.ipac.caltech.edu/forms/calculator.html} of $E(B-V)=0.059$ \citep{Schlegel1998}, we
find  $L_B$ = 1.04 $ \times 10^{11}$~$L_{B,\odot}$. 
For a given galaxy model we can therefore compute the total 
stellar mass-to-light ratio, $M_\star/L_B$, as well as the baryonic fraction that we can compare with 
the observed ones. This requires a choice of IMF. For a Salpeter IMF we 
measured $f_b = 0.65^{+0.13}_{-0.10}$, while for a Kroupa IMF we measured $f_b = 0.45^{+0.04}_{-0.07}$.
In our computation of the baryonic fraction, the total mass is the mass in the Einstein radius, M($<R_E$).

We compared the model properties to the data by computing a chi-square, $\chi_{\rm Jeans}^2$. 
This $\chi_{\rm Jeans}^2$ includes the measured lens velocity dispersion and the total mass in the Einstein radius,
M($<R_E$) = (3.16$\pm$0.31) $ \times 10^{11}$~M$_\odot$. We allowed for a 10\% error on the 
total mass in the Einstein radius to account for the weak dependence of this mass upon 
the choice of a lens model, which led to

\begin{equation}
    \chi^2_{\rm Jeans} = \left(\frac{\sigma_{\rm Jeans} - 222}{34}\right)^2 + \left(\frac{M(<R_E) - 3.16\times 10^{11}}{0.31}\right)^2.
    \label{eq:chijeans}
\end{equation}

\begin{figure}[t!]
\centering
\includegraphics[width=90mm]{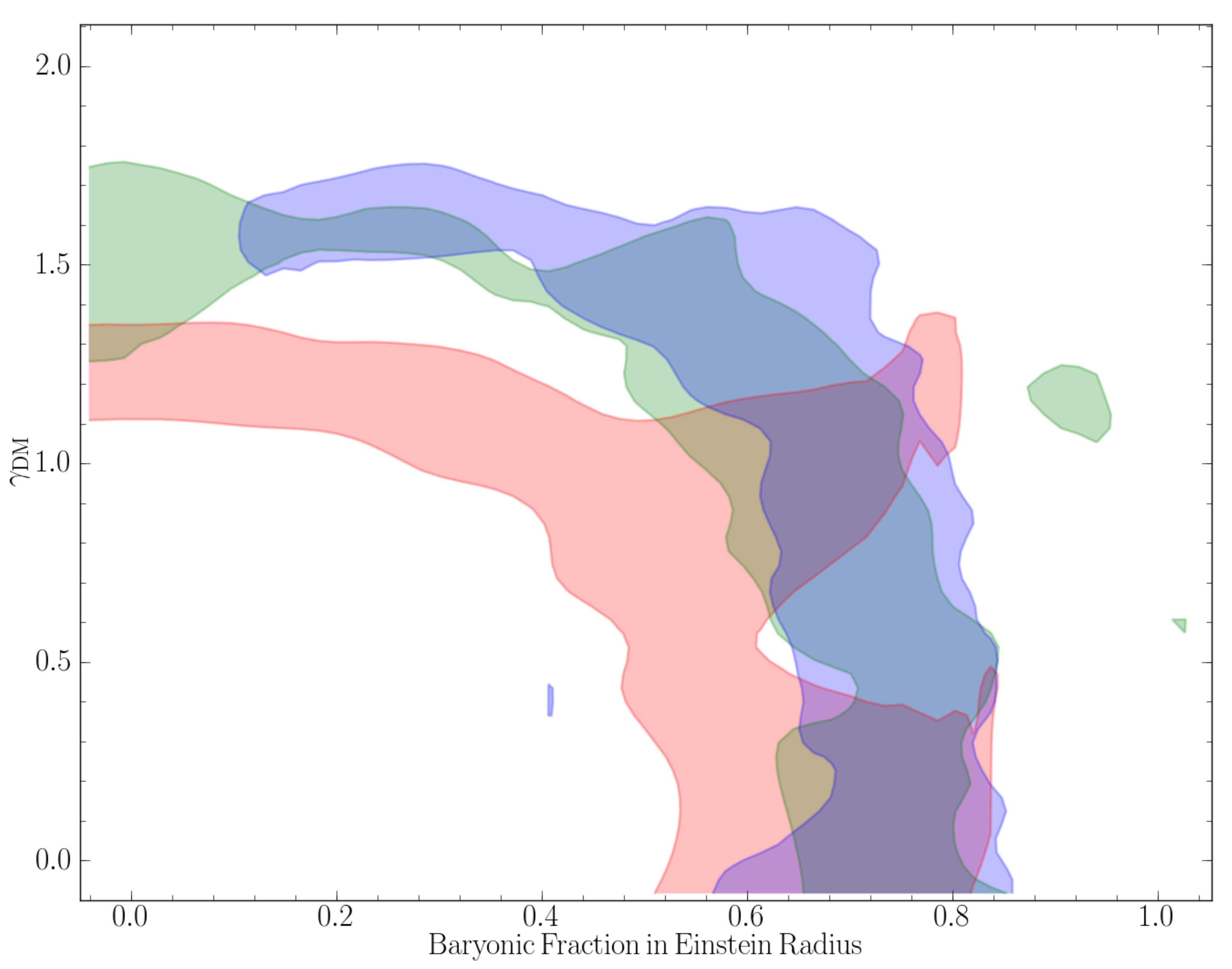}
\caption{Effect of change in $r_s$, the scale radius of the dark matter halo, on the 
$\gamma_{DM} - f_b$ relation. The contours show the region containing 68\% of the models. The blue area
 has $r_s = 25$ kpc (our adopted choice for the final modelling), the green region has $r_s = 15$ kpc,
and the red region is for models with small scale radii, $r_s = 5$ kpc.}
\label{variable_rs}
\end{figure}

\begin{figure}[t!]
\centering
\includegraphics[width=90mm]{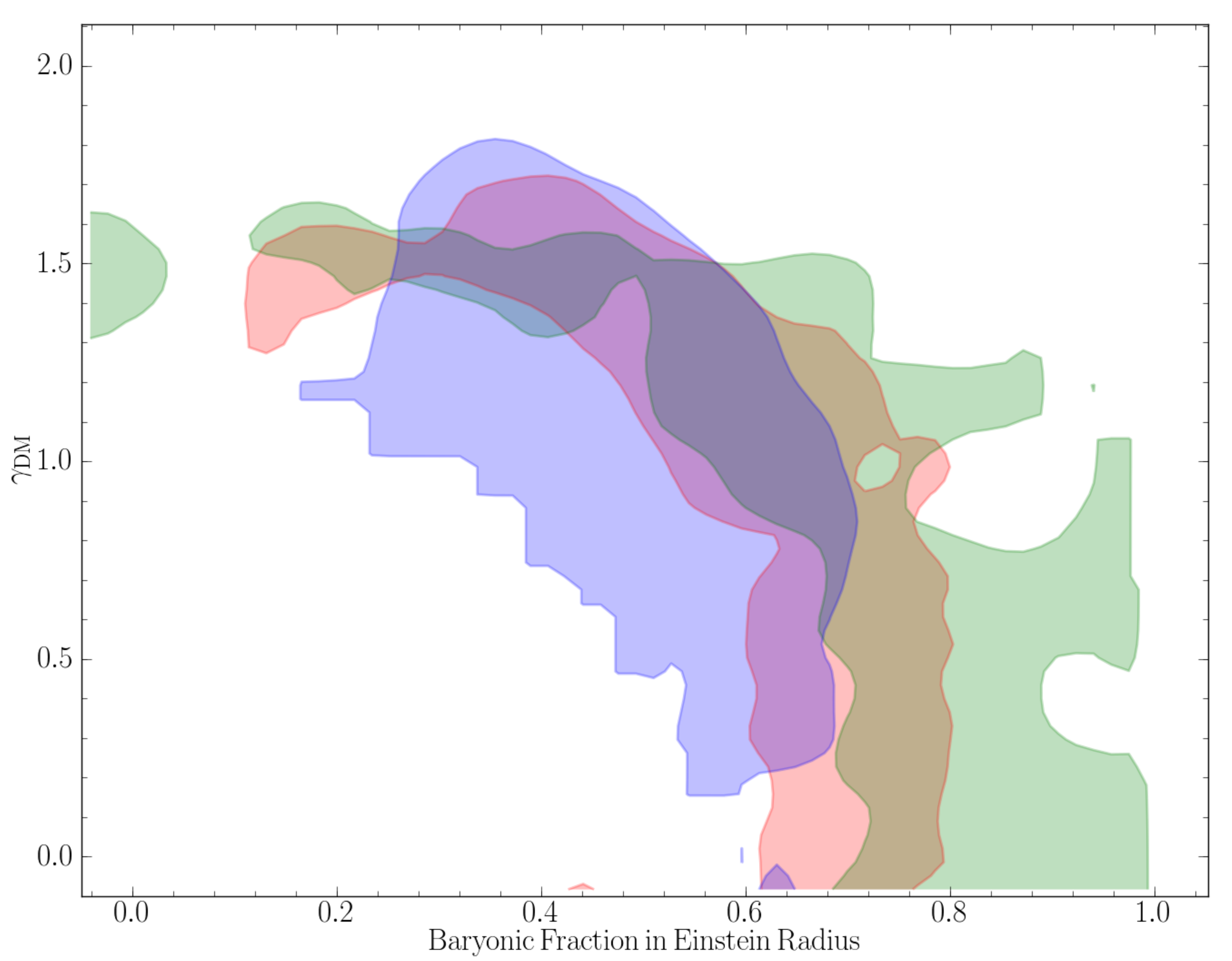}
\caption{Effet of a change in $R_{\rm{eff}}$ on the $\gamma_{DM} - f_b$ relation.  
The contours show the region containing 68\% of the models. The value adopted
in this paper is $R_{\rm{eff}} = 8.44$ kpc, as measured from HST images. This corresponds to the area in red.
The models in the blue region have $R_{\rm{eff}} = 4.22$ kpc, and the region in green has $R_{\rm{eff}} = 12.88$ kpc.}
\label{variable_reff}
\end{figure}

\begin{figure*}[t!]
\centering
\includegraphics[width=85mm, height=75mm]{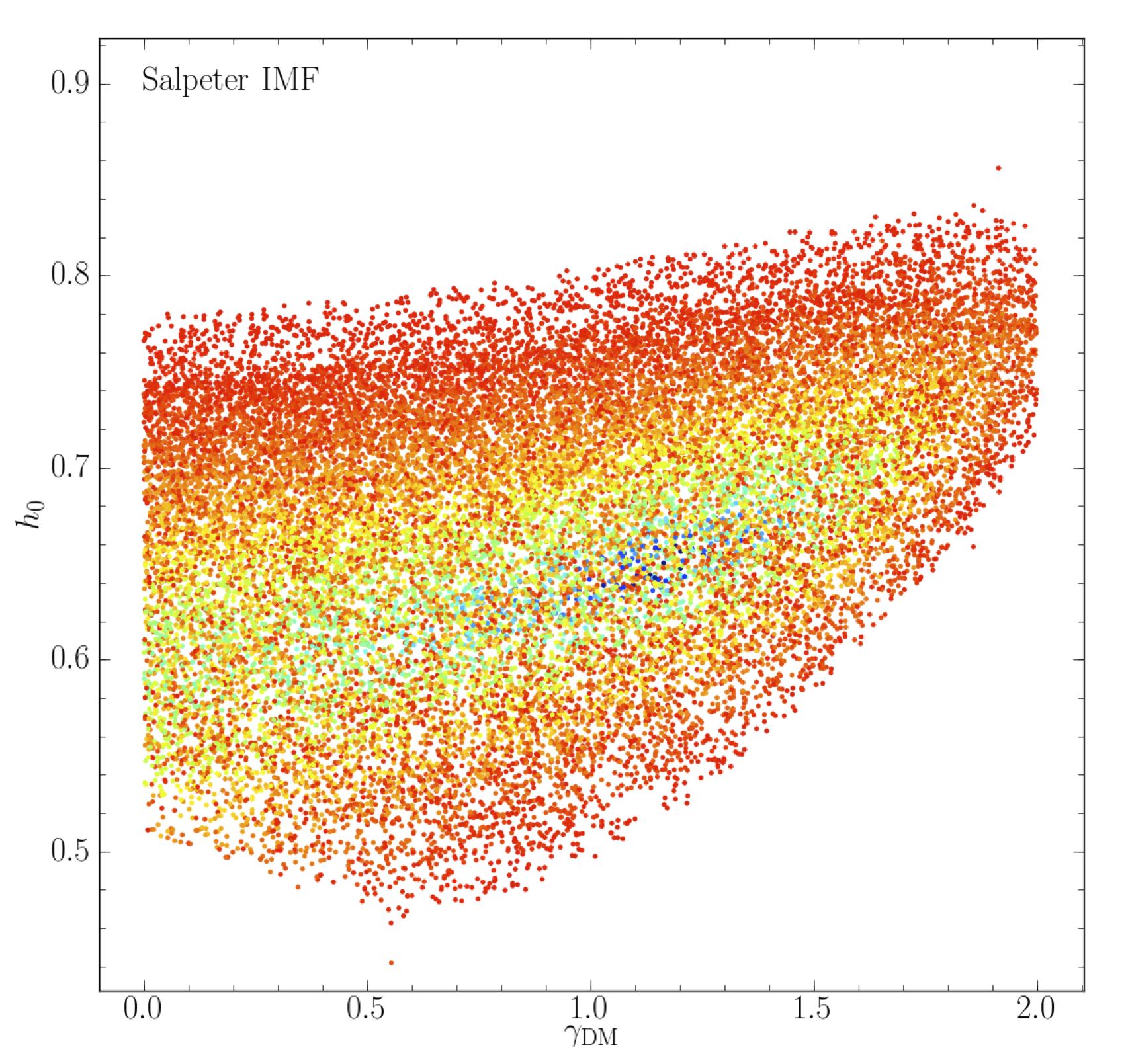}
\includegraphics[width=95mm,height=75mm]{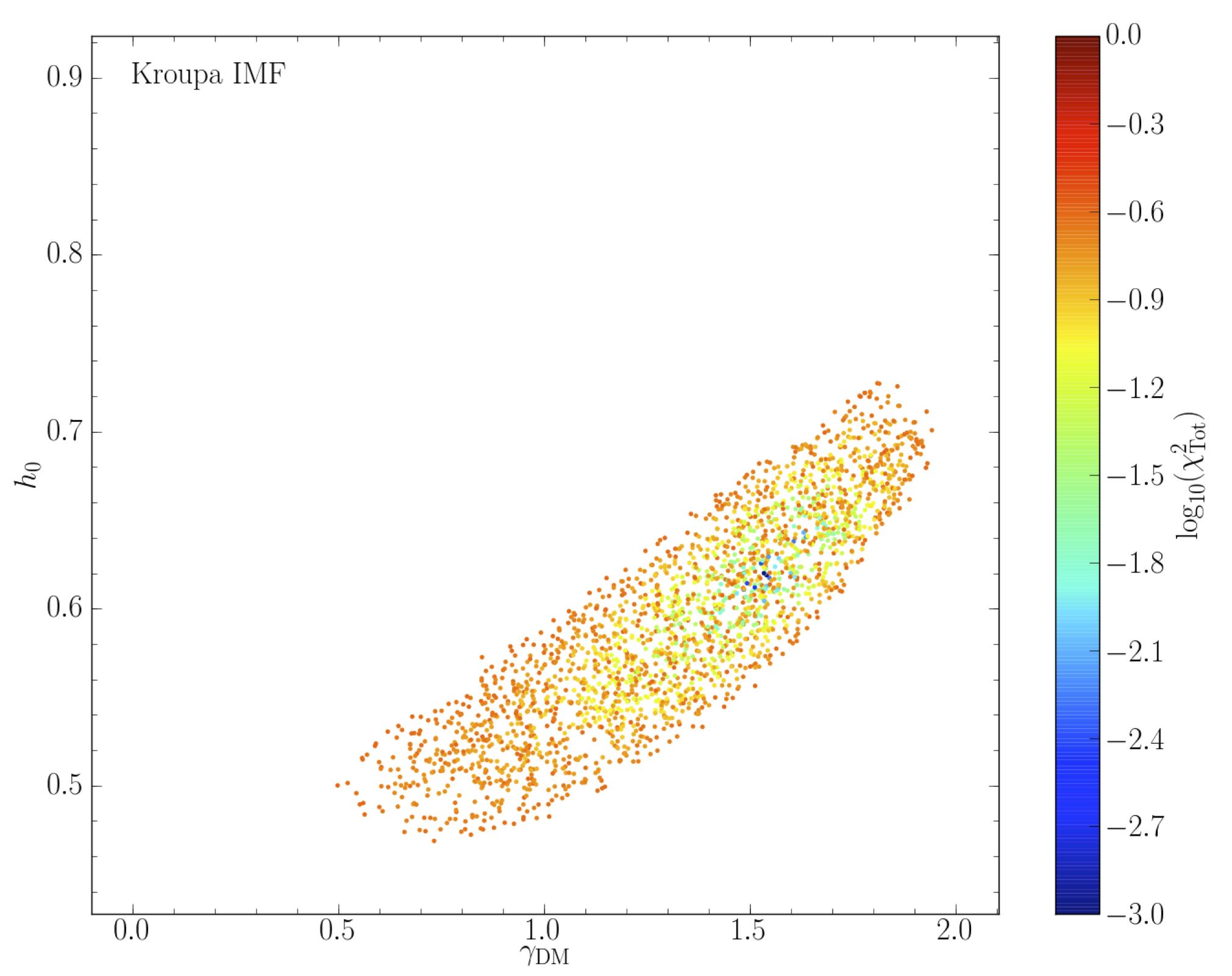}
\caption{Distribution of our lens models as a function of the dark matter slope and \ho. The colour code
gives the value of $\chi^2_{\rm Tot}$ (see text; eq.~\ref{eq:chitot}). Only the models with $\chi^2_{\rm Tot} < 1$ are shown. 
The results in the left panel are for a Salpeter IMF, and
for a Kroupa IMF in the right panel. 
The lensing $\chi^2$ itself is not included in the figure. However, it is systematically lower for galaxies with
Kroupa IMFs than for galaxies with Salpeter IMFs.}
\label{ho}
\end{figure*}

\begin{figure*}[t!]
\centering
\includegraphics[width=85mm, height=75mm]{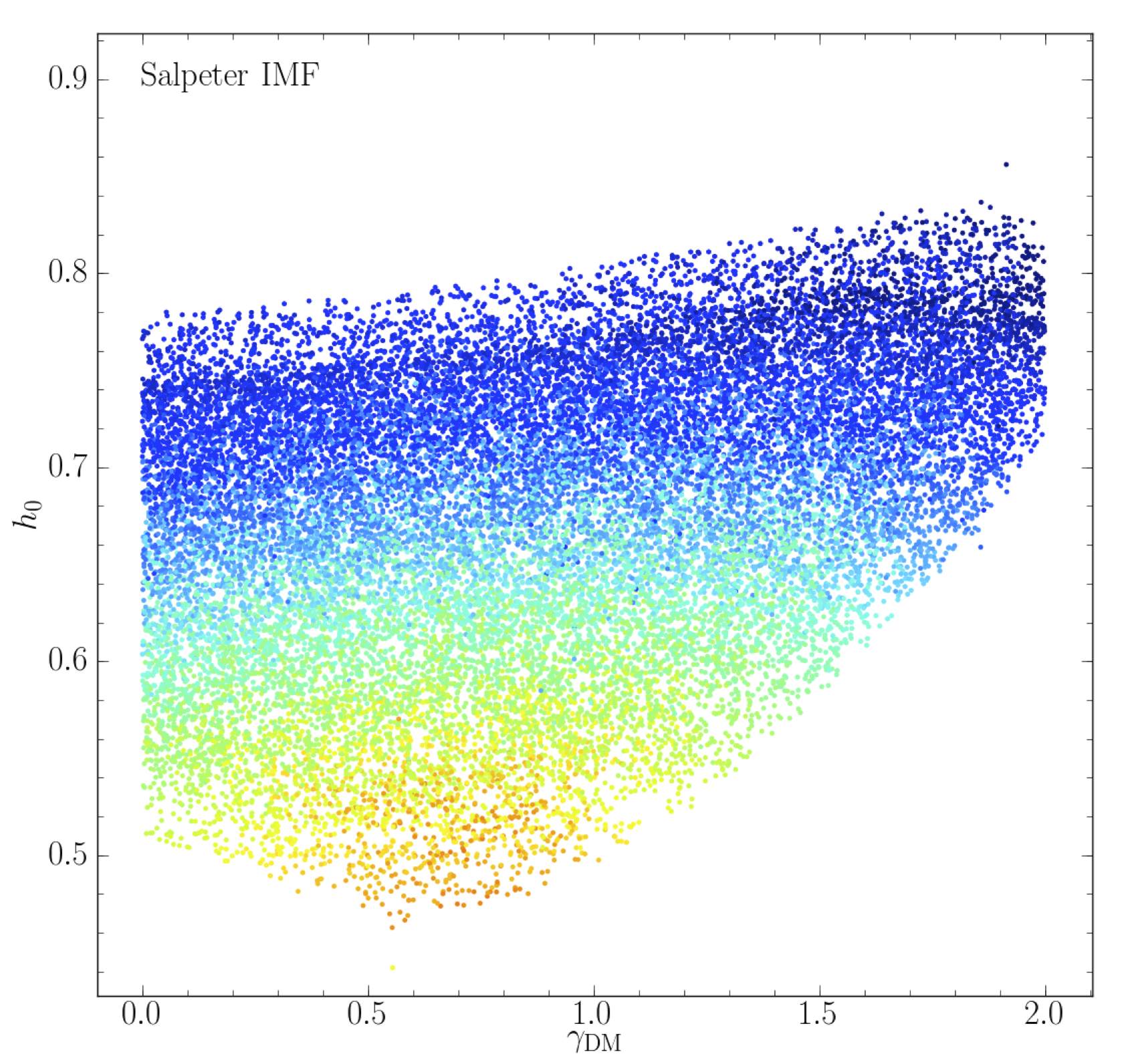}
\includegraphics[width=95mm,height=75mm]{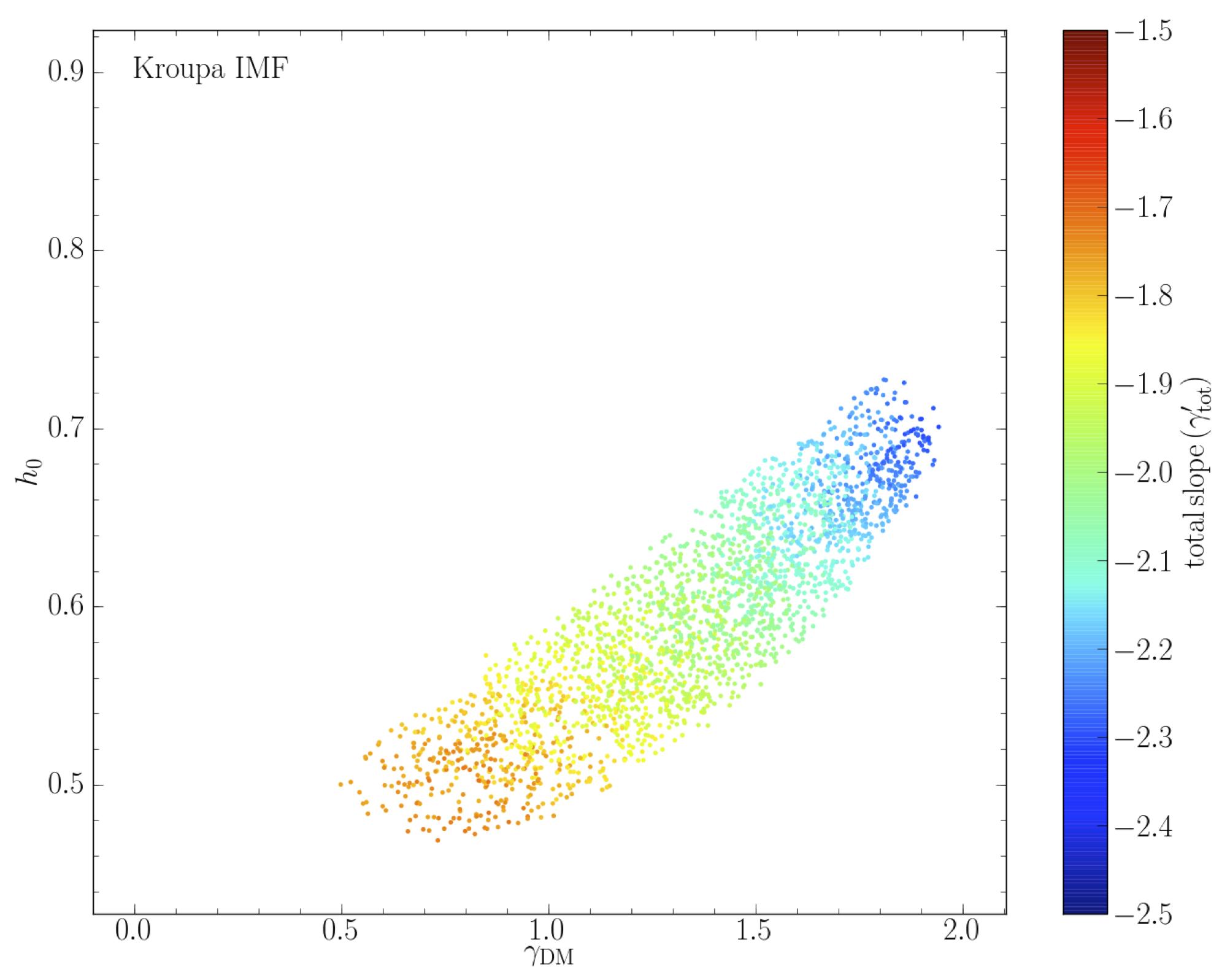}
\caption{Same as Fig.~\ref{ho}, but the colour code now gives the slope of the total (dark+luminous) mass
profile, $\gamma'$. A logarithmic slope of $\gamma'=2$ corresponds to an isothermal profile, shown in green.}
\label{ho-gammatot}
\end{figure*}

We used our numerical integration of the Jeans equations to sample the $\gamma_{DM} - f_b$ plane,
for a given choice of the scale radius, $r_s$, of the dark matter halo. Fig.~\ref{fb-gamma} shows the value 
of $\chi^2_{\rm Jeans}$ for 960 000 galaxy models with $r_s = 4 \times R_E = 25$ kpc.
For a given choice of IMF a precise measurement of the baryonic fraction, $f_b$, 
can constrain the dark matter slope $\gamma_{DM}$ well, at least in the 
case of a Kroupa IMF.

Fig.~\ref{fb-gamma} is drawn for  $r_s = 4 \times R_E = 25$ kpc, which gives the best  $\chi^2_{\rm Jeans}$ values. 
However, models with $\chi^2_{\rm Jeans}<1$ are also obtained with other values of $r_s$. We therefore
test the influence of a change in $r_s$ by building models for a  range of scale radii, 
$r_s = 5, 15, 30$ kpc (i.e., $r_s$ = 0.75$\times$, 2.25$\times$, 4.50 $\times$ $R_E$). 
Fig.~\ref{variable_rs} summarises the results and shows no influence of $r_s$ on the $\gamma_{DM} - f_b$ relation 
unless $r_s$ becomes unrealistically small.

The Hernquist profile in our model also has a scale parameter, $r_\star$, which is fixed to the
observed value of the effective radius, $R_{\rm{eff}} = 8.44$ kpc. Although we have HST images that allow us to measure $R_{\rm{eff}}$ and although our error bars on this parameter are small, we note that 
studies of lensing galaxies by different authors often find different $R_{\rm{eff}}$ values from the same data. This is 
because different methods are used to subtract the quasar images and to mask (or not) the Einstein ring. We therefore
computed models for a range of different $R_{\rm{eff}}$ and show the influence on the $\gamma_{DM} - f_b$ 
plane in Fig.~\ref{variable_reff}, for $R_{\rm{eff}} = 4.22, 8.44, 12.88$ kpc. This reflects an (unlikely high)
50\% change in $R_{\rm{eff}}$ with respect to the actual measurement and translates into a 10-20\% shift
of the $\gamma_{DM} - f_b$ relation. This is comparable to the width of the coloured areas in Fig.~\ref{variable_reff}.
Because our true measurement error on $R_{\rm{eff}}$ is certainly much better than 50\%, we conclude that 
this parameter does not affect the models much, given the accuracy of the observations.  

\section{Towards \ho\, with \obj}
\label{discussion}

The {\tt gravlens} software \citep{Keeton2004} was used 
in combination with the dynamical and stellar population constraints to 
reproduce the lensing configuration of the quasar images and to attempt converting the time
delays into \ho. 
We adopted a total potential well composed of the main lensing galaxy, 
the nearby galaxy G22 \citep[see][]{Morgan2005} plus an external shear. 

The main galaxy is composed of a projected Hernquist + cuspy halo model 
identical to those in eqs. 1 \& 2. 

For the Hernquist profile, we fixed the ellipticity and PA of the lensing galaxy to the observed ones.

We then estimated how well each of the models described in Section~\ref{nbs}
reproduces the observed image configuration. In doing this, 
we allowed only the external shear ($\gamma$, $\theta_\gamma$), the Einstein 
radius of G22 and $H_0$ to vary. The potential well of G22 was assumed to lie at its observed 
position and was modelled as a singular isothermal sphere (SIS). We assumed a conservative value of
$R_E(G22) < 0\farcs4$, following the results of \citet{Morgan2005}, who have found $R_E(G22)=$0\farcs18.

We show in Fig.~\ref{ho} the value of \ho\,  for each
lens model as a function of its dark matter slope, $\gamma_{DM}$.
The colour code in the figure corresponds to the value of $\chi_{\rm Tot}^2$, where the baryonic fraction, $f_b \pm \sigma(f_b)$,
in the Einstein radius is now included in the calculation:

\begin{equation}
    \chi^2_{\rm Tot} = \chi^2_{\rm Jeans} + \left(\frac{f_b ({\rm model}) - f_b ({\rm obs})}{\sigma(f_b)}\right)^2 .
   \label{eq:chitot}
\end{equation}

 Including $f_b$ is justified by 
Fig.~\ref{fb-gamma}, showing that different va\-lues of $f_b$ select different lens models. 
We display our results for the two most common IMFs in use in stellar population
modelling, the Salpeter and the Kroupa IMFs. 

The points define a $\chi_{\rm Tot}^2$ surface with a clear valley which minimum indicates 
the best dark matter slopes for each IMF. These are $\gamma_{DM}({\rm Sal}) \sim 1.15$ and 
$\gamma_{DM}({\rm Kro}) \sim 1.54$ for 
the Salpeter and the Kroupa IMFs, respectively. Each model shown in Fig.~\ref{ho} is also required
to display a good lensing chi-square, $\chi_L^2$, after fitting of the quasar image positions with {\tt gravlens}. 
The values of $\chi_L^2$ are systematically lower for
the lensing galaxies with Kroupa IMFs. The present lensing and dynamical work therefore 
favours a lensing galaxy with a Kroupa IMF, as also found by
\citet{Cappellari2006} from 3D spectroscopy of early-type galaxies and by \citet{Ferreras2008} 
using the SLACS sample of strong lenses \citep{Bolton2008}.

The points in Fig.~\ref{ho} all have $\chi^2_{\rm Jeans} < 1$, which prevents us, for now,
from giving a value for \ho\ given the observational uncertainty on $f_b$ and $\sigma_{\rm ap}$. The
measurement errors on both parameters will, however, easily improve with deeper and higher resolution 
spectroscopy of the lens.

With the current observational constraints we rely on previous work done on the total mass slope,  $\gamma'$,
of lensing galaxies. \citet{Koopmans2009} have measured
the probability distribution function of $\gamma'$ in the 
SLACS sample of strong lenses. They have found $\langle \gamma' \rangle = 2.085^{+0.025}_{-0.018}$,
with an intrinsic spread of $\sigma(\gamma')=0.20$, also confirmed in a more recent study by 
\citet{Auger2010}. The models with the best $\chi_{\rm Tot}^2$ values
in Fig.~\ref{ho} 
correspond to a total slope of $\gamma' = 2.1\pm0.1$ independent of the choice of an IMF. This is well 
within the rms limits of \citet{Koopmans2009} and leads to 57 $<$ \ho $<$ 71 \kmsmpc with
all models equiprobable within this range. 

In our analysis we modelled the environment of the lensing galaxy as a SIS that represents galaxy
G22, plus an external shear with a PA of the order of $-15^{\circ}$, and amplitudes in the range $0.06 < \gamma < 0.08$. 
However, we did not explicitly account for the fact that \obj\, lies within a group of galaxies 
\citep{Momcheva2009}. The unkown convergence, $\kappa$,
associated with the group leads to an overestimate of \ho\, by a factor 
$(1-\kappa)^{-1}$, meaning that the range of acceptable \ho\, values would decrease further if we have 
underestimated the convergence caused by the group. Current imaging data suggest, however, that the line 
of sight is in fact sightly under-dense compared with other lensed systems \citep{Fassnacht2011}. However, the
missing convergence is about $\kappa \sim 0.01-0.02$, leading to at most a 2\% change (upwards) in \ho. 
The above statement should, however, be 
considered with care because the effect of a group is poorly 
appro\-ximated by a simple convergence term. Explicit modelling of the 
group halo is likely needed to properly account for the 
modifications induced on the main lens potential. Deep X-ray and/or optical integral field spectroscopy 
may turn out to be very useful as well in determining the centroid and mass of the group that contains the galaxy lensing 
\obj. 

\section{Conclusion}

We presented seven years of optical monitoring for the four lensed quasar images of \obj. We found 
that the time delays are better expressed with respect to component B, which is the least affected by stellar
microlensing in the lensing galaxy. The formal error bars on the time delays are between 5\% and 
10\% depending on the component, which is remarkable given the very short time delays involved.
In addition, the delays are robust against different tests performed on the data, including removal of 
subsets of data and monte carlo simulations. These tests are possible only with very long light curves,
as provided by COSMOGRAIL. Most past lens monitorings have two to three seasons and much coarser temporal
sampling. The delays are also independent of the way the microlensing
variations are modelled. Given the short length of the delays, additionally improving the error bars will only 
be possible by increasing the temporal sampling of the curves, i.e., by merging all existing data
on \obj\ taken by different groups over the years.  

We introduced a method to convert the time delays into \ho\, purely based on external constraints
on the radial mass slope of the lens. 
These constraints come from deep optical spectroscopy of the lens, which allowed us  to measure its velocity
dispersion and its stellar mass-to-light ratio. The present paper describes our approach,
but the current observations of the lensing galaxy so far lead to a broad range of values for \ho. 

Our methodology complement that presented by \citet{Suyu2009, Suyu2010} well. 
Combining the two in the future will require follow-up observations such as (i) deep 
HST imaging to map the Einstein ring with high signal-to-noise, (ii) deep high resolution spectroscopy 
of the lens over a broad spectral range to narrow down the uncertainty on 
its velocity dispersion and stellar mass, (iii) integral field spectroscopy 
to measure all redshifts within 30-60\arcsec\ around the lens, and (iv) X-ray imaging to pinpoint 
massive groups along the line of sight.
 
If all these observations are taken for \obj\ as well as for a few lenses with well measured time delays, the cost in 
terms of follow-up observations will still remain very modest compared with other existing methods to measure \ho,
such as Cepheids and supernovae.

\begin{acknowledgements}
We are grateful to  all the observers who contributed to the data acquisition at the Euler and Mercator telescopes 
as well as at Maidanak Observatory. COSMOGRAIL is financially supported by the Swiss National Science Foundation (SNSF). This work is also supported by the Belgian Federal Science Policy (BELSPO) in the framework of the PRODEX Experiment Arrangement C-90312. VC thanks the Belgian National Fund for Scientific Research (FNRS). DS acknowledges 
a fellowship from the Alexander von Humboldt Foundation. MK has been supported by the Programa Nacional de 
Astronom\'{\i}a y Astrof\'{\i}sica of the Spanish Ministry of Science and Innovation under grant
\emph{AYA2007-67752-C03-01} and DO02-85/2008 from Bulgarian Scientific Research Fund.
\end{acknowledgements}

\bibliographystyle{aa}
\bibliography{HE0435_revised}
\end{document}